\documentclass[fleqn,usenatbib]{appmnras}
\usepackage{newtxtext,newtxmath} 
\usepackage[whole]{bxcjkjatype}
\usepackage[T1]{fontenc} 
\DeclareRobustCommand{\VAN}[3]{#2}
\let\VANthebibliography\thebibliography
\def\thebibliography{\DeclareRobustCommand{\VAN}[3]{##3}\VANthebibliography}
\usepackage{graphicx}	
\usepackage{amsmath}	
\usepackage{bm}  
\usepackage{color}   
\usepackage{mfirstuc}
\def\be{\begin{equation}}\def\ee{\end{equation}}
\def\vlsr{v_{\rm LSR}}  \def\vr{v_{\rm r}}
\def\deg{^\circ}  \def\vlsr{v_{\rm lsr}} \def\vrot{V_{\rm rot}}
\def\vlsr{V_{\rm lsr}} \def\Vrot{V_{\rm rot}}
\def\co{$^{12}$CO }  
 
\def\Xco{X_{\rm CO}}   \def\Tb{T_{\rm B}}

 \def\kms{km s$^{-1}$}   
   
 \def\htwo{H$_2$} \def\Tb{T_{\rm B}}

 \def\apj{ApJ} \def\aap{AA} \def\mnras{MNRAS} \def\pasj{PASJ} 
 \def\araa{ARAA}\def\apjl{ApJ.L.}
 \def\xcounit{H$_2$ cm $^{-2}$ [K km s$^{-1}]^{-1}$}

\def\vrot{V_{\rm rot}}  \def\ve{V_{\rm e}} \def\vr{v_{\rm r}}

\def\aaps{A\&ApS}
\def\mnras{MNRAS}

\title[Face-on Map of the Molecular Disc and 3-kpc Expanding Ring]{Face-on Map of the Molecular Disc and 3-kpc Expanding Ring of the Galaxy based on a High-Accuracy Rotation Curve}

 \author[Y. Sofue]{Yoshiaki Sofue \\ 
Institute of Astronomy, The University of Tokyo, Mitaka, Tokyo 186-0015, Japan}
\date{Accepted; Received YYY; in original form} 
\pubyear{2023}  

\begin{document} 
\maketitle     

\begin{abstract}
We analyze the longitude-velocity diagram (LVD) of \co\-line emission from archival data and use the most accurate rotation curve (RC) of the Milky Way to transform radial velocity to face-on position in the galactic plane.
We point out that the face-on transformation is highly sensitive to the adopted RC, especially in the inner Milky Way, in the sense that deviations of the RC from the true rotation velocity lead either to an artifact hole or overcrowded concentration along the tangent circle for over- or under-estimated RC, respectively.
Even if the RC is sufficiently accurate, non-circular motion such as with the 3 kpc expanding ring introduces significant artifacts in the resulting face-on-map, as long as a circular rotation is assumed.
On the other hand, if we properly take into account the non-circular motion, it can be used to solve the near-far degeneracy problem of determination of kinematic distance.
We thus propose a new method to solve the degeneracy by incorporating the expanding motion of a ring or arms.
We apply the method to the LVD of the 3-kpc expanding ring and present its face-on map projected onto the galactic plane for the first time.
\end{abstract}


\section{Introduction} 

The 3-kpc expanding ring of interstellar gas in the Milky Way is recognized as a tilted elliptical feature in the longitude-velocity diagram (LVD) of the HI and CO line emissions in the Galactic plane
\citep{1972A&A....16..118S,1974ApJ...188..489S,1977ARA&A..15..295O,2008ApJ...683L.143D}.
The oval LV feature has been attributed to arm or ring structure formed either by an expanding motion of a shock front produced by an explosive event in the Galactic Center
\citep{1974ApJ...188..489S,1976PASJ...28...19S,1977A&A....60..327S}
or by non-circular motion of gas in an oval potential of a stellar bar
\citep{1999A&A...345..787F}.
Due to the highly non-circular velocities, the kinematic method to determine the distance by transformation of the radial velocity cannot be applied to this particular structure.

Since the pioneering work by \cite{1958MNRAS.118..379O}, there have been extensive studies for mapping the face-on structure of the Milky Way using the radial velocities of the HI and molecular gases on the assumption of circular rotation (hereafter, "face-on transformation, or FOT") 
\citep{1986A&AS...65..427B,2003PASJ...55..191N,2006PASJ...58..847N,2007A&A...469..511K,2016PASJ...68....5N,2016PASJ...68...63S}. 
Because these works have aimed at mapping the global spiral structure in the entire Galaxy, the inner arm structures have not necessarily been resolved well, although the molecular map can trace the 4-kpc molecular ring \citep{2016PASJ...68....5N,2016PASJ...68...63S}.

Recently, \cite{2023PASJ...75..279F} have analyzed the molecular gas distribution in the inner Galaxy between $l=10\deg$ and $62\deg$ using the high-resolution \co-line data from FUGIN (Four-receiver system Unbiased Galactic plane Imaging survey with the Nobeyama 45-m telescope)\citep{2017PASJ...69...78U}.
By applying a machine-learning near-far distinguishing method of molecular clouds, they have obtained a high-resolution face-on map of the molecular gas using FOT for a flat circular rotation.
They have identified several inner spiral arms, and noticed an artificial hole in the central region.
Besides the identified arms, an arm-like structure is evident in their face-on map, which apparently extends from the Scutum arm, crosses the far-side solar circle at $l\sim 12\deg$, and extends beyond the circle, composing a massive "leading" spiral structure.

In this paper, we seek for the reason to produce such irregular structures often appearing in the current face-on maps of the Milky Way by considering the non-circular motion associated with the 3-kpc expanding arm. 
We further derive a face-on distribution of the molecular gas in the Galactic plane by analyzing the archival \co\-line data. 

\begin{figure} 
\begin{center}      
\includegraphics[width=8cm]{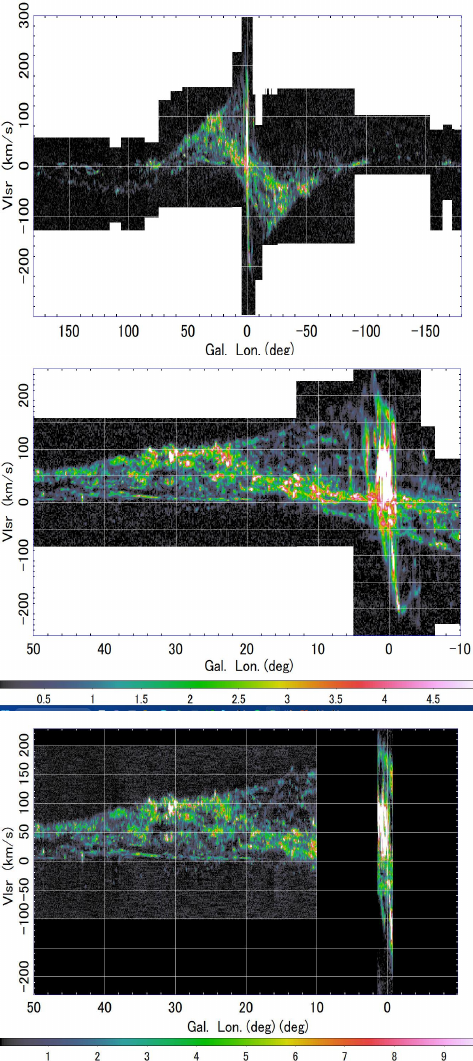}
\end{center}
\caption{[Top] LVD of \co-line $\Tb$ in K at $b=0\deg$ from Columbia survey \citep{2001ApJ...547..792D}.
{[Middle] Same, but from $l=-10\deg$ to $+50\deg$.}
[Bottom] LVD from FUGIN with 45-m telescope in the same longitude range as above  \citep{2017PASJ...69...78U} and GC surveys \citep{2019PASJ...71S..19T}.
All diagrams are smoothed, not presenting the original resolutions.
} 
\label{fig-lvd1}  	
\end{figure}

\section{Longitude-velocity diagrams}

We make use of the archival \co\-line data cubes from the Columbia and FUGIN surveys.
{The radial velocity $\vlsr$ used in this paper is the LSR (local standard of rest) velocity.}
The LVD of the entire disc is taken from the Columbia survey \citep{2001ApJ...547..792D}, which had angular resolution of $9'$ and velocity coverage of $\pm 300$ \kms, and the used LVD at $b=0\deg$ had $2881\times 493$ pixel sizes in the longitude and velocity directions, which was resized to a $1000\times 1000$ pixel sized LVD, having $0\deg.36$ and 0.6 \kms pixel sizes.

LVDs from FUGIN surveys integrated between $b=-1\deg$ and $+1\deg$ have been published, in which various known galactic structures and molecular clouds have been identified \citep {2019PASJ...71S...2T,2021PASJ...73S.129K}.
An LVD along the Galactic plane at $b=0\deg$ has been published by \cite{2021PASJ...73L..19S} in order to derive the rotation curve.
In this paper, we use FOT to cover the entire galactic disk, including the distant region outside the solar circle $\sim 20$ kpc away.
Since the integrated LVD in the $b$ direction weights near and far clouds equally, not only does the face-on map give more weight to the far side disc, but also the treated disc thickness as a function of distance also increases.
To avoid this inconvenience, and for the convenience of data processing, we use the LVD at $b = 0\deg$ here.
This eliminates the need to consider the height distribution of the gas, but does not allow us to discuss the 3D structure of the disk.

The full beam width at half maximum of the 45-m telescope was $15''$ at the \co ($J=1-0$)-line frequency, and the velocity coverage and resolution were $\pm 250$ \kms and 1.3 \kms, respectively, and the rms noise levels were $\sim 1$ K.  
The original pixel size was $(\Delta l, \Delta b, \Delta \vlsr) =$ (8''.5, 8''.5, 0.65 \kms).
The FUGIN LVD is then combined with that in the Galactic Centre between $l=-1\deg$ and $1\deg.4$ with the same angular and spectral resolutions from the GC CO line survey \citep{2019PASJ...71S..19T}.
The thus combined 45-m LVD was resized to $1000\times 1000$ pixel sized LVD between $l=-10\deg$ and $+50\deg$, having $0\deg.06$ and 0.5 \kms pixel sizes.
Thus obtained LVDs are shown in Fig. \ref{fig-lvd1}, and are used individually to obtain face-on maps using FOT.

The resulting face-on maps from the three data sets (Columbia, FUGIN and GC) will be merged on the $(X,Y)$ plane, where $X$ and $Y$ are the Cartesian coordinate axes with the origin at the Galactic Centre and are positive toward $l=90\deg$ direction and toward the Sun, respectively.
The distance to the Galactic center from the Sun ($R_0$) is assumed to be $R_0=8.0$ kpc \citep{2020PASJ...72...50V,2019ApJ...885..131R}.

\section{Rotation curve}

The rotation curve (RC) is the most important and sensitive parameter for the conversion of radial velocity to kinematic distance.
As mentioned in the previous section, the current studies have used fairly simple analytical models that approximate the overall behavior of the observed rotation velocities.
While acceptable, thought not accurate enough, in the mid to outer disc, the currently adopted rotation curves do not necessarily represent the Galactic rotation with sufficient accuracy in the inner disc and the Galactic Centre (GC) region. 

Deviations between the assumed RC and the actual rotation speed will artificially increase or decrease the calculated density.
If the model RC is underestimated compared to the true velocity, FOT produces an overestimated density near the tangent circle.
This occurs because gas observed at higher velocities than the model collects gas from the surrounding region, and, furthermore, gas that does not satisfy the equation is "abandoned" (forgotten) from solution.
This means that the total mass of the gas is not conserved during the FOT.

Conversely, if the RC is overestimated, the gas is "swept" away from the right position to the surrounding area and the density is underestimated.
As the consequence it creates an artificial "hole" on the tangent circle.
In fact, a flat rotation curve fixed at $V=240$ \kms, about $R\sim 40$ \kms faster than the actual rotation at $R\sim 2-4$ kpc, has yielded an artificial hole along the tangent circle \citep{2023PASJ...75..279F}.

\begin{figure} \begin{center}   
\includegraphics[width=8cm]{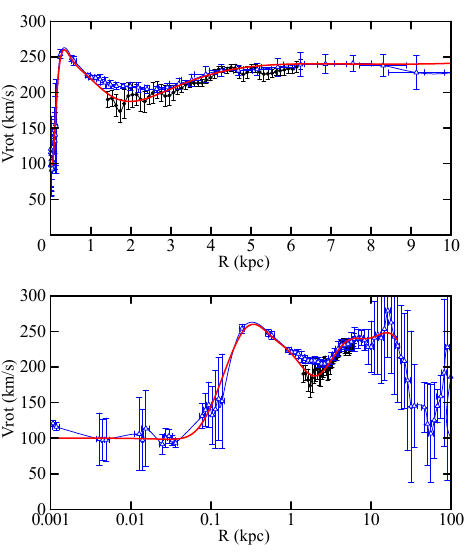}   
\end{center}
\caption{Rotation curves from FUGIN (black dots: \citep{2021PASJ...73L..19S}), from compilation for the entire Galaxy from the nucleus to the halo (triangles: \citep{2013PASJ...65..118S}, and a model curve by the red line used for the distance determination which approximately fits the observations. } 
\label{fig-rc}  	\end{figure}

To avoid or minimize such artifacts, we adopt here the most accurate rotation curve obtained so far for the Milky Way.
Figure \ref{fig-rc} shows the rotation curve, where the black dots represent the internal rotation curve obtained by terminal velocity fitting using the FUGIN CO-line LVD \citep{2021PASJ...73L..19S}. 
Triangles are taken from the grand rotation curve from the central black hole of the GC to the galactic outskirts, which was constructed by compiling currently published RC data and fitting to the central CO line LVDs \citep{2013PASJ...65..118S}. 
The red line shows the empirically fitted curve adopted for the FOT in this paper.

\section{Face-on map for circular rotation by FOT-C}

\begin{figure} 
\begin{center}    
\includegraphics[width=8cm]{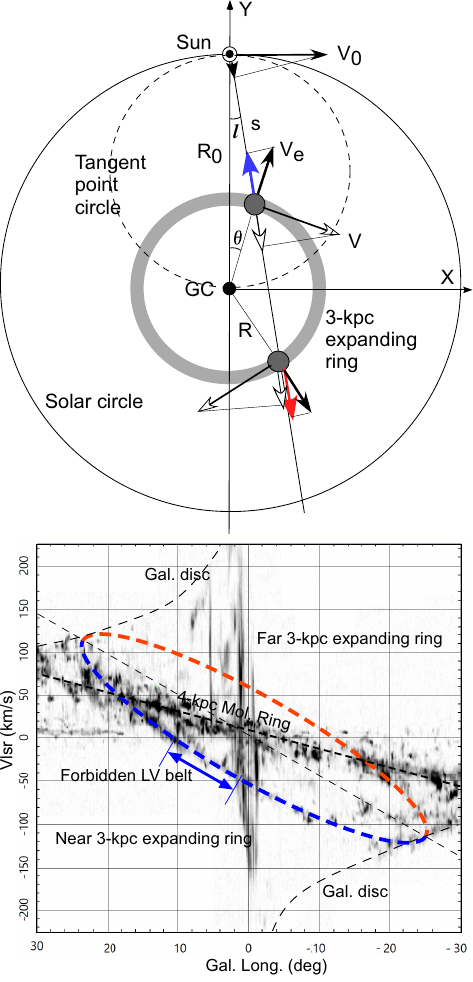}  
\end{center}
\caption{Illustration of the velocities in rotation and expansion, and
LVD from Columbia survey \citep{2001ApJ...547..792D} with explanation of the structures discussed in this paper.} 
\label{fig-illust}  	
\end{figure}

We first apply the face-on map transformation assuming the circular rotation, which we call the FOT-C method.
The radial velocity is related to the rotation velocity by
\begin{equation}
    \vr=\left( \vrot \frac{R_0}{R}-V_0\right) \sin\ l.
    \label{eq-vrcirc}
\end{equation}
The distance $s$ of the object from the Sun is given by
\begin{equation}
    s=R_0 \cos\ l \pm \sqrt{R^2-R_0 ^2 \sin ^2 l}.
    \label{eq-s}
\end{equation}
The $+$ and $-$ signs stand for the far- and near-side solutions, respectively.
At the distance $s$, the volume density of the \htwo\ gas is calculated by
\begin{equation}
    n_{\rm H_2}=dN_{\rm H_2}/ds=\Xco\Tb dv/ds,
\end{equation}
where $N_{\rm H_2}=\Xco\int \Tb dv$, $\Tb$ is the \co\ brightness temperature, $\Xco=2\times 10^{20}$ \xcounit is the CO-to-\htwo\ conversion factor,  and the velocity gradient along the line of sight, $dv/ds$, is calculated by
\begin{equation}
    dv/ds=\sqrt{(dv_1/ds)^2 + (dv_2/ds)^2},
\end{equation}
where 
\begin{equation}
    \frac{dv_1}{ds}= 
    \left| \frac{R_0}{R}\left(\frac{d\vrot}{dR} - \frac{\vrot}{R} \right)\sin\ l ~ \cos\ (l+\theta) \right|
\end{equation}
is the gradient due to Galactic rotation, and
\begin{equation}
    dv_2/ds=v_\sigma/\Delta s \sim 5 \ {\rm km\ s^{-1} kpc^{-1}}
\end{equation}
is the internal gradient due to turbulent motion of the gas of the order of $v_\sigma\sim 5$ \kms which gives the minimum value of gradient along each line of sight.
Here,  $R$ and $R_0=8$ kpc are the Galacto-centric radius and solar distance from the GC, respectively, $l$ and $\theta$ are the Galactic longitude and Galacto-centric longitude, $\vrot$ and $V_0=238$ \kms are the rotation velocity at $R$ and $R_0$, as explained in Fig. \ref{fig-illust}.

\begin{figure*} 
\begin{center}      
\includegraphics[width=15cm]{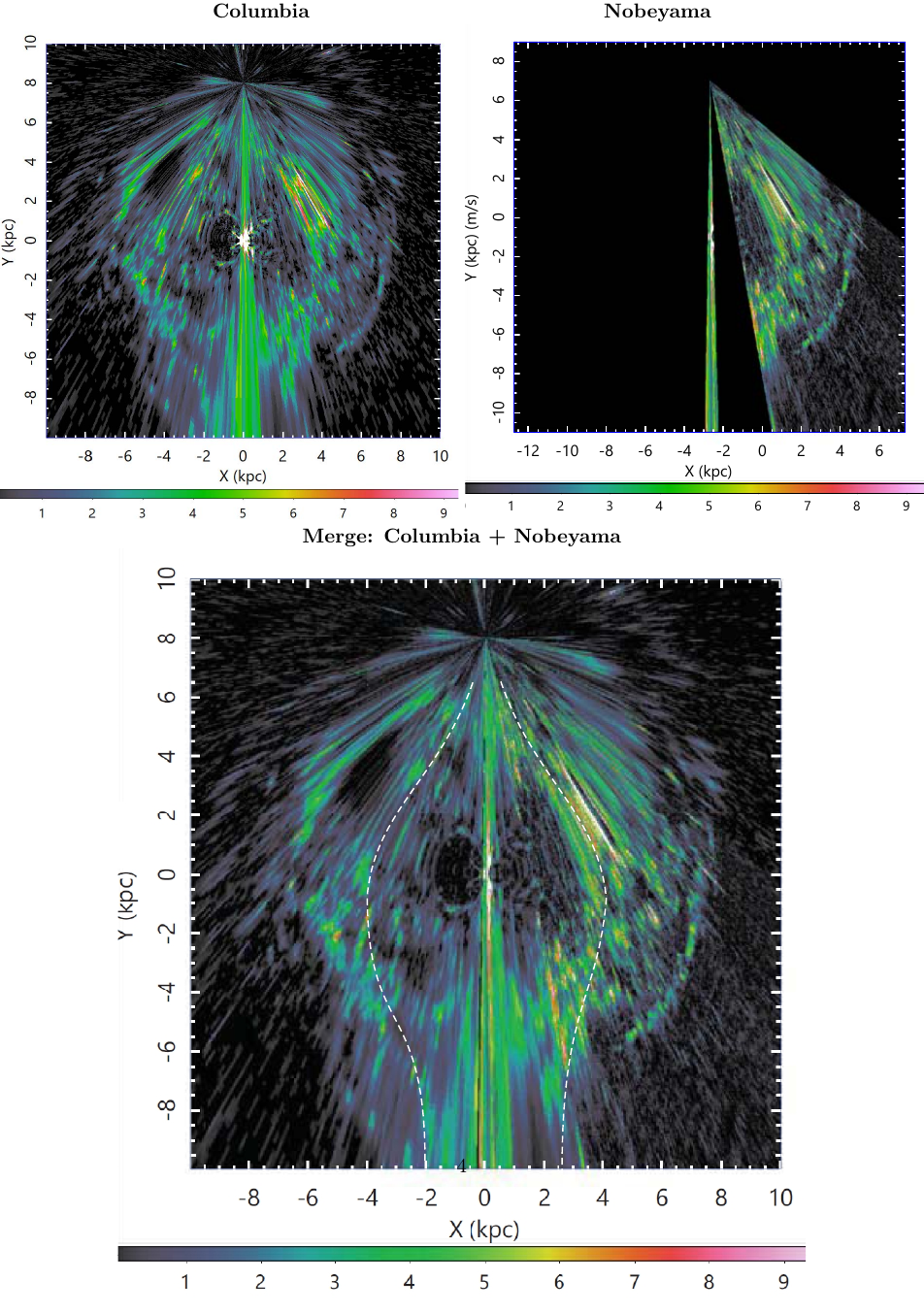}   
\end{center}
\caption{[Top left] FOT-C maps of \htwo\ density in \htwo\ cm$^{-3}$ from the Columbia and [top right] Nobeyama CO-line LV diagrams for circular rotation. 
Bottom panel shows composite, where the region of panel (a) which is observed at Nobeyama has been replaced with map (b).
Near and far emissions are duplicated in these maps. 
The white dashed lines trace artifact "bows" corresponding to the 3-kpc expanding ring caused by the assumption of circular rotation. 
The far-side massive "leading" arm is due to the (erroneous) location of the "forbidden" LV belt (Fig. \ref{fig-illust}), which is actually the near-side expanding (approaching) ring.
}   
\label{fig-faceon}  	
\end{figure*}

Figure \ref{fig-faceon} shows the face-on maps of the volume density of \htwo\ gas obtained by FOT-C using the rotation curve in Fig. \ref{fig-rc} and LVDs in Fig. \ref{fig-lvd1} under an assumption of pure circular rotation.
The results from the both surveys are merged according to their longitudinal coverage.   
The near and far emissions are duplicated in this map, so that the map is not appropriate to discuss such non-axisymmetric structure like spiral arms.
The circular assumption results in such a duplicated artifact of the nearby features such as the sharp ring-like arm along the far-side solar circle, running coherently from $l\sim 20\deg$ to $\sim 50\deg$, which is the consequence of erroneous location of the structure associated with the Aquila Rift. 
However, it still provides with some basic information about the axi-symmetric structures such as the 4-kpc molecular ring and some arms as identified in the earlier works \citep{2016ApJ...823...77R}.

Due to the high accuracy of the rotation curve in the 1st and 4th quadrant (northern disc of the Milky Way), the artifact "hole" does not appear any more on the tangent circle in the right side of the GC in this map.
However, hole-like regions still remains in the left side in the 2nd and 3rd quadrants (southern disc).
Such lopsidedness of the holes are due to the asymmetric rotation curves between the northern and southern Milky Way, which are represented here by a single RC model for convenience. 

Another notable artifact feature in this map is the two long bow-shaped arms symmetric about the $Y$ axis, as indicated by the two dashed white lines.
The positive $X$ side bow runs towards $l\sim 20\deg$ near the Sun, crosses the 4 kpc molecular ring at $(X,Y)\sim (2,+4) $ kpc, returns almost parallel to the $Y$ axis, crosses the 4kpc ring in the far side again at $(3,-4)$ kpc, and further extends to $(2,-9)$ kpc, finally crossing the distant solar circle.
This bow feature is also visible in the map by \cite{2023PASJ...75..279F} as a massive "leading arm" across the circle beyond the solar circle.
The bow on the negative $X$ side is also traced symmetrically.

Such a bow structure extending for $\sim 10-20$ kpc across the inner disc and the solar circle is difficult to explain by any astrophysical mechanism in the Galactic disc.  
The bow cannot be removed even using the sophisticated near-far deconstruction procedure, as demonstrated in the recent detailed study using the Nobeyama high-resolution data \citep{2023PASJ...75..279F}.
In the next section, we argue that the bow structures are artifacts due to the assumption of circular rotation applied to the 3-kpc expanding ring in a highly non-circular motion. 

We also mention that the narrow fan region around the $Y$ axis (Sun-GC line) filled with straight features extending from the Sun in the GC direction is composed mainly of artifacts of strongly deformed central molecular zone (CMZ) due to the insufficient resolution and too much crowded equal-velocity lines (Fig. \ref{fig-vfield}). 
So, this central fan region is out of the analysis hereafter.

\section{Expanding ring with near-far separation by FOT-Ex}

\begin{figure} 
\begin{center}   
(a)\includegraphics[width=8cm]{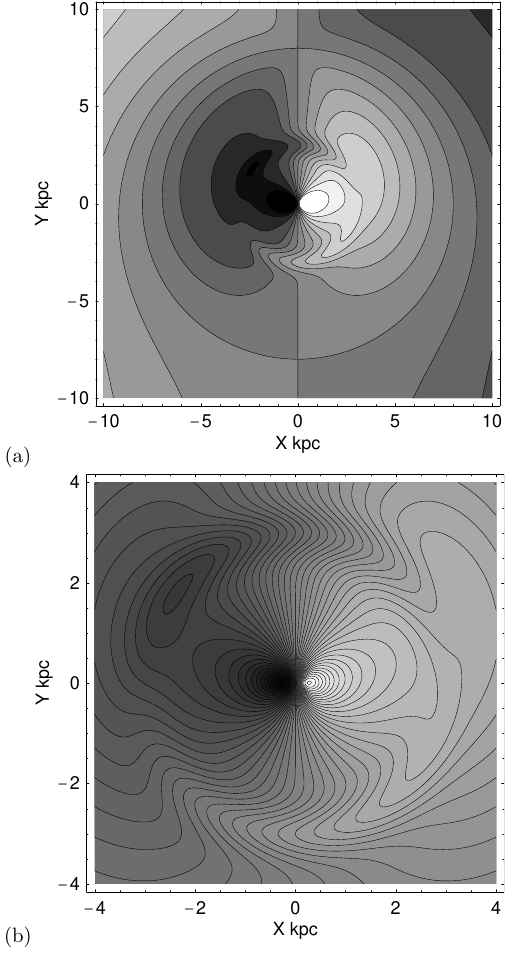}    
\end{center}
\caption{(a) Radial velocity field of the Milky Way with an expanding ring of a radius 3 kpc,  and (b) close up in the inner region. The complicated velocity field demonstrates that the radial-velocity to space transformation on an assumption of simple circular rotation results in a considerably deformed map from the true distribution.} 
\label{fig-vfield}  	
\end{figure}

In addition to the accuracy of rotation curve, non-circular motion also affects the face-on transformation.
In this section we propose a new method to solve the degeneracy of near-far distances incorporating the expanding motion, which we call the FOT-Ex method.
The radial velocity $\vr$ superposed by an expanding motion of $\ve$ (positive outward)
is given by
\begin{equation}
    \vr=\left( \vrot \frac{R_0}{R}-V_0\right) \sin\ l \pm \ve ^0 \cos\ (l+\theta).
    \label{eq-vrex}
\end{equation}
In Fig. \ref{fig-vfield} we show the distribution of $\vr$, or a velocity field, in the Galactic plane in the $(X,Y)$ coordinates (Fig. \ref{fig-illust}) as calculated for a flat rotation curve superposed by an expanding ring of radius 3 kpc, width 1 kpc, and expanding velocity $\ve=50$ \kms. 
The velocity field is significantly deformed from the general symmetric butterfly pattern with respect to the Sun-GC line ($Y$ axis).

The complex velocity field shows that FOT-C, assuming simple circular rotation, leads to highly deformed maps that are quite different from the actual distribution.
This complex behavior of the velocity field offers a unique opportunity to resolve the degeneracy of near-far solutions on the same line of sight.
However, it is impractical to transform the measured LVD into a face-on map without knowing the object's motion, whether it follows a pure circular rotation or a combination of circular and extensional motions.
So we perform a separation on the LV diagram, where the 3 kpc ring is recognized and traced as a tilted LV ellipse as in Fig. \ref{fig-illust}.

\begin{figure} 
\begin{center}    
\includegraphics[width=6.2cm]{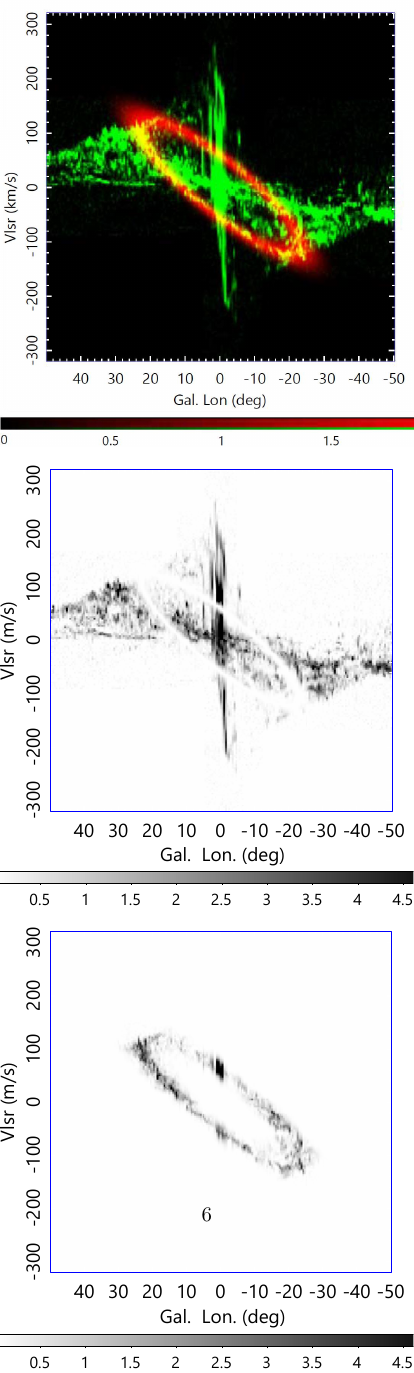}   
\end{center}
\caption{[Top] LVD at $b=0\deg$ from Columbia CO survey \citep{2001ApJ...547..792D} at longitude within $\pm 50\deg$ (green) superposed by a tilted ellipse representing the 3-kpc expanding ring (red).
[Middle] Same, but the LV ellipse has been removed.
[Bottom] Same, but only the elliptical region representing the 3-kpc ring.} 
\label{fig-lvd}  	
\end{figure}

The 3-kpc expanding ring shows up as a tilted ellipse in the LV diagrams in Fig \ref{fig-lvd} as reproduced from the Columbia CO survey \cite{2001ApJ...547..792D}.
Panel (a) shows the ellipse by a red oval, and (b) shows the LVD after subtracting the emission along the ellipse. Panel (c) shows an LVD of the subtracted elliptical component. 
  
\begin{figure} 
\begin{center}        
\includegraphics[width=6.5cm]{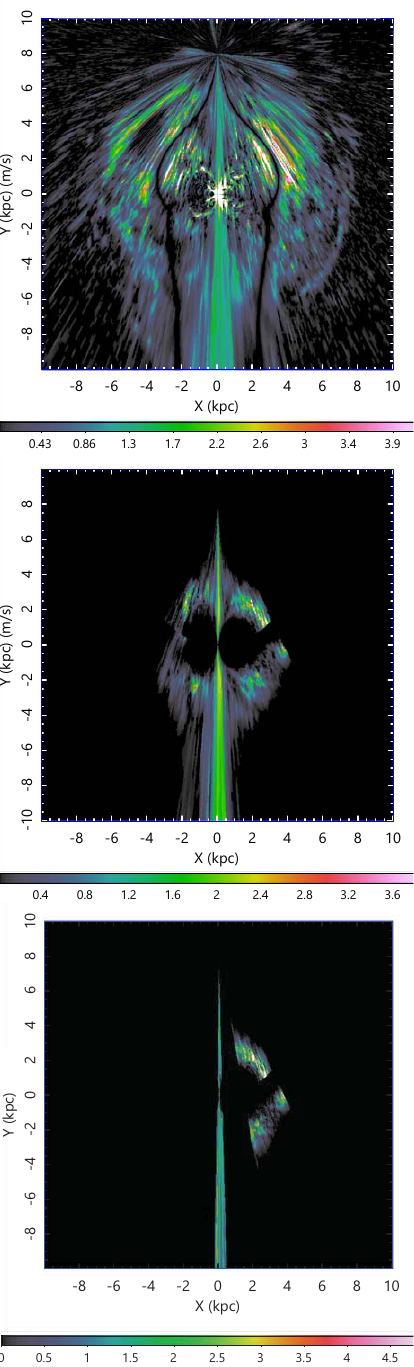}   
\end{center} 
\caption{[Top] Face-on projection of $n_{\rm H_2}$, where the LV ellipse representing the 3-kpc expanding ring (Fig. \ref{fig-lvd}) has been removed.
[Middle] 3-kpc expanding ring corresponding to LV ellipse in Fig. \ref{fig-lvd} middle panel.
Near-far degeneracy has been resolved because of the expanding motion.
[Bottom] Same, but using Nobeyama 45-m data.
}   
\label{fig-ring}  	
\end{figure}

We first apply the simple FOT-C with circular rotation to the LVD in Fig. \ref{fig-lvd}(b) using Eq. (\ref{eq-vrcirc}), where the ellipse component (3 kpc ring) has been subtracted.
The result is shown in Fig \ref{fig-ring}(a), where the dark bows represent the eliminated disc gas corresponding to the LV ellipse caused by the 3-kpc expanding ring.

\begin{figure} 
\begin{center}       
\includegraphics[width=8cm]{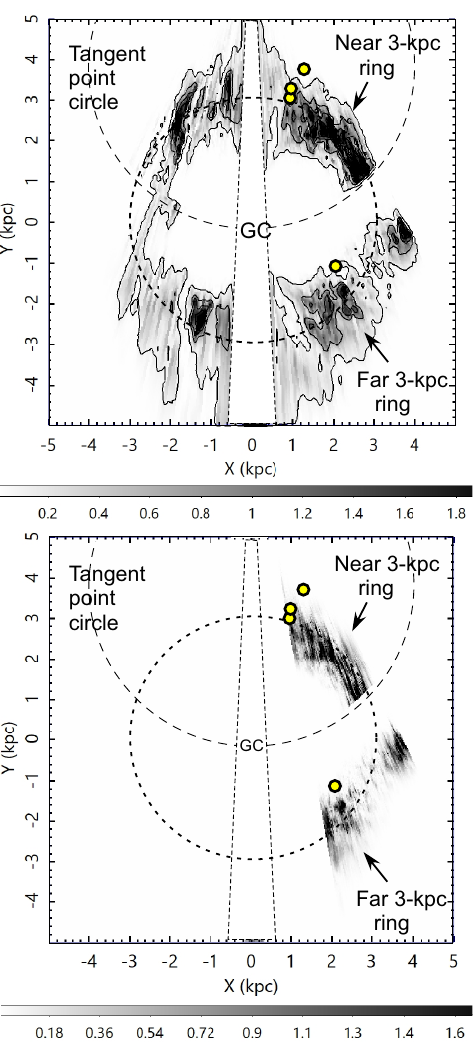}     
\end{center}
\caption{
[Top] Close up of the 3-kpc expanding ring from Columbia survey. The clear-cut gap on the tangent point circle is artifact. The triangular region near the $Y$ axis (Sun-GC line) is ambiguous due to the $\vlsr$ degeneracy and is avoided from the analysis. 
{Yellow circles are maser sources associated with the 3-kpc expanding arms with known trigonometric distances \citep{2019ApJ...885..131R}, which lye on a circle slightly ($\sim 0.5$ kpc) displaced from the molecular ring to the direction of the Sun.}
[Bottom] Same, but using the 45-m CO data. 
}   
\label{fig-ring-up}  	
\label{fig9}
\end{figure}

We then apply FOT-Ex using Eq. (\ref{eq-vrex}) to the 3 kpc ring component represented by the tilted LV ellipse in Fig. \ref{fig-lvd}(c). 
This solves the degeneracy problem by choosing the corresponding side of the ellipse depending on the approaching (near) and the receding (far) arms.
The results are shown in Fig. \ref{fig-ring}(b) for the Columbia data and in (c) for the Nobeyama data.
Figure \ref{fig-ring-up} enlarges the resulting face-on maps of the 3 kpc ring and adds descriptions of the visible components.
Considering the radial motion of the gas on the expanding ring, the emission of the lower half of the tilted LV ellipse is assigned to the near side in FOT-Ex, and the upper half to the far side.
This separation eliminates near-far degeneracy.
On the other hand, the resulting map shows a clear cut along the tangent circles, as seen in the resulting map as indicated by the dashed half circle in Fig. \ref{fig-ring-up}.

\section{{Discussion}}

\subsection{{Accuracy of the maps}}
{The accuracy of the FOT-C method has been discussed in 
detail by \cite{2011PASJ...63..813S}, where the rotation curve has been assumed to be valid, while the measurement of the radial velocity includes errors.
In the present study, the accuracy of velocity measurement of individual molecular clouds is on the order of the resolution of the \co\ spectrometer, $\sim 1$ \kms, sufficiently small compared to the observed velocity $\vlsr$ of $\sim 100$ \kms.}
 
\def\ds{\delta s}
\def\dVrot{\delta \Vrot}
\def\Vexp{V_{\rm exp}^0}
\def\dVexp{\delta \Vexp}
\def\sin{{\rm sin}}
\def\cos{{\rm cos}}
\def\sinl{\sin\ l}

{
The uncertainty of the FOT map in this study arises mainly from the uncertainty of the assumed rotation velocity $\Vrot$ and expansion velocity of the 3-kpc ring $\Vexp$.
We here examine the effects of these quantities, $\dVrot$ and $\dVexp$, respectively, on the distance determination whose error (uncertainty) is denoted by $\ds$.
Using equations \ref{eq-vrex} and \ref{eq-s}, the uncertainties propagate to $\ds$ as follows\footnote{{Small deviation of a multi variable function $f(x_i)$ due to small deviation $\delta x_i$ is assumed to be expressed by $\delta f^2 \simeq \Sigma_i \left(\frac{\partial f}{\partial x_i}\delta x_i\right)^2$ }}.
\be
\ds\simeq A \left[\left(\frac{\dVrot}{B}\right)^2+\left(\frac{\Vrot\dVexp}{B^2}\right)
^2 \right]^{1/2}, 
\label{eq-ds}
\ee
where
\be 
A=\frac{R \ \sinl}{\sqrt{(R/R_0)^2-\sin^2 l}}
\label{eq-A}
\ee 
($R\ge R_0 \sin \ l$) and
\be
B=\vr - V_0 \ \sinl \pm \Vexp \cos(l+\theta).
\label{eq-B}
\ee  
 }

{Equation \ref{eq-A} indicates that the error (uncertainty) attains maximum along the tangent-point circle with $R=R_0 \sinl$, along which the result is most sensitive to the rotation curve error.
On the other hand, the effect of the rotation curve is minimized as $\propto \sinl$ near the Sun-GC line, where the radial velocity is degenerated to zero.
}

{ 
We then estimate a typical value of distance error (uncertainty) in the FOT map in a representative region around $R\sim 3$ kpc, $l\sim 20\deg$, and $\vr\sim 80$ \kms, assuming that the rotation curve has uncertainty of $\delta \Vrot = 10$ \kms, as read from Fig. \ref{fig-rc}.}

{First, we consider a case of purely circular rotation, so that the second term of Eq. \ref{eq-ds} is ignored, or $\ds \sim A \delta \Vrot/B$.
Then we obtain $\delta s\sim 0.41$ kpc, which is proportional to the uncertainty $\delta \Vrot$.
{If the rotation curve is sufficiently accurate with $\delta \Vrot \lesssim 5$ \kms, which is the case in the present study in the 1st and 2nd quadrants ($l=0\deg$ to $90\deg$) of the disc, where the RC determination was made.
The distance uncertainty is only $\ds \sim 0.2$ kpc here, and the FOT map is reasonably accurate.
However, in the 3rd and 4th quadrants ($l=270\deg$ to $360\deg$), the RC may not be accurate enough, which causes the hole-like dark region in the left side of the GC and around $(X,Y)\sim (-4, +3)$ kpc in Fig. \ref{fig-faceon}.}}

If we adopt a flat rotation curve which overestimates the rotation velocity by $\delta \Vrot \sim +30$, the distance error is as large as $\ds \sim 1.3$ \kms, yielding significant under/over estimation of the distance in the near/far side.
This is the reason for the artifact 'hole' in the FOT map in the inner disc obtained for the flat rotation curve \citep{2023PASJ...75..279F}, where the gas has been swept away from the right position for $\pm 1.3$ kpc, making a hole of diameter $\sim 2.6$ kpc.

We next consider the case including the expanding ring, which is assumed to have expanding velocity of $\Vexp=50$ \kms.
The distance uncertainty is estimated to be $\ds\sim 0.55$ for $\delta \Vexp\sim 2$ \kms, and $0.76$ kpc for $\sim 4$, for the same uncertainty of rotation curve $\delta \Vrot = 10$ \kms as above, while it varies according to Eq. \ref{eq-ds}.

{To summarize, the uncertainty of the distance determination by FOT-Ex is about $\sim 0.5$ kpc for $\delta \Vrot\sim 10$ \kms around the 3-kpc ring. 
It increases toward the tangent circle attaining the maximum there, and is minimum along the Sun-GC line.
However, it must be recalled that the distance determination has a maximum uncertainty along the Sun-GC line for another reason that the radial velocity is degenerated to zero, so that the error due to $\delta \vlsr$ increases to infinity \citep{2021PASJ...73L..19S}.
For this reason, the plot near the Sun-GC line ($l\sim 0\deg$) is avoided in Fig. \ref{fig-ring-up}.}

\subsection{{Comparison with trigonometric parallax distances of star forming regions associated with the 3-kpc expanding arms}}

{There have been detected a number of maser sources along the tilted ellipse on the LVD corresponding the 3-kpc expanding ring \citep{2011ApJ...733...27G}.}
It is reported that the trigonometric parallaxes of maser sources associated with the star-forming region G9.62+0.20 correspond to a distance of $5.2\pm 0.6$ kpc, while its LSR velocity is $\sim 2$ \kms  \citep{2009ApJ...706..464S}.
This velocity corresponds to kinematic distances of 0.5 or 16 kpc, if a circular rotation is assumed, which contradicts the trigonometric distance.
The trigonometirc distance is consistent with the here derived distance of the molecular 3-kpc arm at $l\sim 10\deg$ in Fig. \ref{fig-ring-up}, where the LSR velocity is $\sim 2$ \kms.
Therefore, we may reasonably consider that G9.62+0.20 is associated with the 3-kpc expanding molecular ring.

Several more maser sources with trigonometric distances are located at $R\sim 3$ kpc ($s\sim 5$ kpc) associated with the near 3-kpc expanding arm, and one source near the far-side arm \citep{2019ApJ...885..131R}, as plotted by yellow circles in Fig. \ref{fig9}.   
For their circular alignment aong the 3-kpc ring the maser sources are considered to be physically associated with the molecular ring.

However, a closer look at the figure suggests that the maser sources are systematically displaced from the molecular ring toward the Sun by about $\sim 0.5$ kpc, whish is also demonstrated by an off-center circle fitted to the sources by \cite{2019ApJ...885..131R}.
If we rely on the maser parallax distances and consider that the sources are physically associated with the 3-kpc molecular ring, we need a modification of the ring model, so that it produces an off-center ring and asymmetric expansion.
In fact, a slightly lopsided expansion is observed as the faster expanding velocity of the far side ring at $\sim 60$ \kms \citep{2008ApJ...683L.143D} than the near side ring at 50 \kms.
{However, this pauses a puzzling problem why the galacto-centric radius of the far-side arm is smaller ($\sim 2.5$ kpc) than the near side arm ($\sim 3.2-3.5$ kpc) despite the faster expansion in the far side. }
 
{\subsection{Expanding ring vs non-circular motion by a bar} }

{The present result does not exclude the possibility that the expanding "ring-like" feature on the LVD is due to a non-circular motion induced by the oval orbits in a barred potential 
\citep{1991MNRAS.252..210B,1999ApJ...522..699A,2022ApJ...925...71L}. 
The FOT-Ex method assumes that the 3-kpc ring is expanding at $\sim 50$ \kms. If the same amount of radial motion can be assumed, any model, such as the bar model, would be able to lead to a similar face-on map as here.
However, the nearly perfect elliptical nature of the 3-kpc ring on the LVD without nodal crossing near $l\sim 0\deg$ seems to be in favor of the expanding ring model.
The bar potential model predicts that the arms on the LVD have either nodes around $l\sim 0$ drawing a tilted 8 shape (not oval) or a parallelogram structure, as often observed in barred spiral galaxies including the 4-kpc molecular ring of the Milky Way.
In anyway, the present analysis is not accurate enough to distinguish a ring from an ellipse. 
Namely, we cannot exclude the possibility that the 3-kpc expanding "ring" is due to a bar potential, and, vise versa, the bar theory cannot rule out the expanding ring model in the present accuracy. }

\vskip 2mm
\section{Summary}

We have shown that face-on transformations (FOT) of molecular-line radial velocities yield maps showing circular structures representing molecular arms when suitable rotation curves are used.
However, when the model RC deviates from the true rotational velocity, an artifact hole along the tangent circle occurs for an overestimated RC, and anomalous gas concentrations for underestimated RC.

Even if the RC is sufficiently accurate, the non-circular motion due to the 3 kpc expanding ring causes considerable artifacts in the resulting face-on-map as long as circular rotation is assumed.
Such artifacts are seen as giant leading arms across the far side of the solar circle.
This is a result of the erroneous positioning of the nearby 3 kpc expanding ring, which includes the "forbidden velocity" belt with negative velocity at $l\sim 0\deg$ to $+12\deg$ as shown in Fig. \ref{fig-illust}.
The approaching LV ridge is transformed to a near-side 3kpc ring when we extract the lower half of the tilted elliptical LV component and apply FOT-Ex considering the expanding motion.  
The heavy leading arm on the far side, which resulted from the assumption of circular rotation, has disappeared in this FOT-Ex map.
The 3-kpc ring on the far-side \citep{2008ApJ...683L.143D} is also clearly visible on the face-on map as the opposite ring beyond the tangent circle by the transformation of the upper half of the LV ellipse.

Finally, we comment on the limitations and future prospects of this study.
This paper is the first to point out, mainly through the analysis of large-scale LVDs in the Columbia survey, that the problem of near-far degeneracy around the tangent circle can be solved by using non-circular motions of the arm and ring.
The LVD from Nobeyama's high-resolution observations has been smoothed to an angular resolution of $0\deg.06$, so the original resolution ($20''$) is not properly incorporated into the current analysis.
The Nobeyama CO LVD at the Galactic Centre is also significantly smoothed and thus did not provide useful information about gas distribution in the CMZ.
Since we used LVD only in the galactic plane at $b=0\deg$, we were unable to touch on the vertical (3D) structure of the disk and ring.
Higher-resolution and 3D analyses of the Nobeyama LVDs inside the Milky Way galaxy as well as in CMZ, which was avoided here as the artifact fan region, would be a subject for future work. 

\section*{Acknowledgments}
The author is indebted to the FUGIN team (Prof. T. Umemoto, et al.) for the CO data observed with the Nobeyama 45-m telescope operated by the NAOJ (National Astronomical Observatory of Japan).
The data analysis was carried out on the computer system at the Astronomy Data Center of the National Astronomical Observatory of Japan.  
{The author is indebted to the anonymous referee for the valuable comments to improve the paper. }

\section*{Data availability}
The FUGIN data were retrieved from the JVO portal at {http://jvo.nao.ac.jp/portal}.

\section*{Conflict of interest}
The author declares that there is no conflict of interest.

{}   

\end{document}